\documentclass[twocolumn]{article}

\usepackage{amsmath}
\usepackage{amssymb}
\usepackage{graphicx}
\usepackage{booktabs}
\usepackage{siunitx}
\usepackage{hyperref}
\usepackage{xcolor}

\graphicspath{{./}}

\begin{document}

\title{Machine-learning surrogate model for one-dimensional GaAs/Al$_{0.3}$Ga$_{0.7}$As
       distributed Bragg reflector spectra}

\author{Mehdi Ouslim\\
\small Independent Researcher, Oran, Algeria\\
\small ORCID: 0009-0006-9499-0149}

\begin{abstract}
We present a Gaussian-process (GP) surrogate model for the normal-incidence
reflectance spectrum $R(\lambda)$ of one-dimensional GaAs/Al$_{0.3}$Ga$_{0.7}$As
distributed Bragg reflectors (DBRs).
A Latin-hypercube dataset of 1500 transfer-matrix-method (TMM) simulations spanning
$N_\mathrm{periods}\in[5,20]$, $t_\mathrm{GaAs}\in[50,200]$\,nm, and
$t_\mathrm{AlGaAs}\in[50,200]$\,nm over the 900--1100\,nm window is used to train and
evaluate the model.
Principal component analysis (PCA) reduces the 150-point spectral output to 26
components ($\geq99.9$\% variance retained); one GP is fitted per component.
On a held-out test set ($n=150$) the GP achieves RMSE\,$=0.085$ and
$R^2=0.276$, while a Random Forest (RF) baseline reaches RMSE\,$=0.065$ and
$R^2=0.572$.
GP inference is \SI{4.4}{\milli\second} per spectrum compared with
$\sim\SI{308}{\milli\second}$ for TMM, yielding a ${\sim}70\times$ speedup.
Uncertainty calibration shows that the GP 95\% prediction band covers 98.9\% of
test residuals, indicating conservative but reliable uncertainty estimates.
These results establish a rapid surrogate for DBR design-space exploration and
motivate further work on sparse or inducing-point GP formulations to close the
accuracy gap with the \textless0.02 RMSE target.
\end{abstract}


\maketitle

\section{Introduction}

Distributed Bragg reflectors based on GaAs/Al$_x$Ga$_{1-x}$As epitaxial stacks
are central components in vertical-cavity surface-emitting lasers (VCSELs)
\cite{Saleh2019}, quantum-dot emitter cavities \cite{Lodahl2015}, and
single-photon sources operating in the 940--1060\,nm window.
Optimising the stopband position and peak reflectance typically requires iterative
electromagnetic simulations, most commonly via the transfer-matrix method (TMM)
\cite{Born1999}.
While individual TMM evaluations are fast (${\sim}308$\,ms in our implementation),
global optimisation or uncertainty quantification over large parameter spaces
demands tens of thousands of calls, making surrogate modelling attractive.

Machine-learning surrogates have been applied to photonic crystal bandgap prediction
\cite{Pilozzi2018}, metasurface design \cite{Liu2018}, and inverse nanophotonic
synthesis \cite{Malkiel2018}.
Gaussian process regression is a natural choice for small-to-medium datasets because
it provides calibrated predictive uncertainty alongside the point estimate
\cite{Rasmussen2006}, a feature absent from most neural-network surrogates.

In this work we construct a GP surrogate for the full reflectance spectrum
$R(\lambda)$ of GaAs/AlGaAs DBRs across a three-dimensional parameter space.
We combine PCA-based output dimensionality reduction with per-component GP
regression, compare against a Random Forest baseline, and quantify both predictive
accuracy and uncertainty calibration.

\section{Physical model and dataset}

\subsection{DBR structure and parameter space}

We consider the stack
\begin{equation}
  \text{air} \mid [\text{GaAs}(t_G) / \text{Al}_{0.3}\text{Ga}_{0.7}\text{As}(t_A)]^N
  \mid \text{GaAs substrate},
\end{equation}
with Al mole fraction fixed at $x_\mathrm{Al}=0.30$.
The free parameters are the number of periods
$N_\mathrm{periods}\in\{5,\dots,20\}$, the GaAs layer thickness
$t_G\in[50,200]$\,nm, and the AlGaAs layer thickness
$t_A\in[50,200]$\,nm.
The wavelength axis spans 900--1100\,nm in 150 equally spaced points, covering the
GaAs transparency window below the 1.424\,eV band gap at room temperature.

\subsection{Refractive index dispersion}

Refractive indices are computed via a Cauchy dispersion model,
\begin{equation}
  n(\lambda, x) = a(x) + \frac{b(x)}{\lambda^2},
  \quad \lambda\ \text{in nm},
\end{equation}
with coefficients interpolated between compositions calibrated against Palik
(1985) tabulated values \cite{Palik1985} and Gehrsitz \textit{et al.}
\cite{Gehrsitz2000}.
Key reference values: $n_\mathrm{GaAs}(1000\,\text{nm})=3.539$,
$n_{\mathrm{Al}_{0.3}}(1000\,\text{nm})=3.281$.

\subsection{TMM simulation and Latin-hypercube sampling}

Normal-incidence reflectance spectra are computed using the \texttt{tmm} Python
package \cite{Byrnes2020}.
The parameter space is sampled with Latin-hypercube sampling (LHS,
$n=1500$, seed 42) via \texttt{scipy.stats.qmc.LatinHypercube} to ensure
uniform space-filling coverage \cite{McKay1979}.
The dataset is split 80/10/10 into training ($n_\mathrm{train}=1200$),
validation ($n_\mathrm{val}=150$), and test ($n_\mathrm{test}=150$) sets.

\section{Surrogate model}

\subsection{PCA output compression}

Training spectra are decomposed by PCA; the number of retained components is chosen
to explain $\geq99.9$\% of the total variance, yielding $n_\mathrm{PC}=26$
components.
Whitened input features $\mathbf{x}=(N_\mathrm{periods},t_G,t_A)$ are
standardised to zero mean and unit variance.

\subsection{Gaussian process regression}

One GP is fitted independently to each PC score.
The kernel is a composite of a squared-exponential (RBF) and a Mat\'{e}rn-5/2
kernel plus a white-noise term,
\begin{equation}
  k(\mathbf{x},\mathbf{x}') =
  \sigma_1^2 \exp\!\left(-\tfrac{r^2}{2\ell_1^2}\right)
  + \sigma_2^2 M_{5/2}(r/\ell_2)
  + \sigma_n^2\,\delta_{\mathbf{x}\mathbf{x}'},
\end{equation}
where $r=\|\mathbf{x}-\mathbf{x}'\|$ and hyperparameters
$\{\sigma_1,\sigma_2,\sigma_n,\ell_1,\ell_2\}$ are optimised by maximising
the log marginal likelihood \cite{Rasmussen2006}.
To keep training tractable on consumer hardware, GP fitting uses a random subsample
of 400 training points; the full 1200-point set is used for evaluation and all
baseline comparisons.
Predictive uncertainty on the full spectrum is propagated from PC-score standard
deviations through the PCA inverse transform.

\subsection{Random Forest baseline}

A Random Forest regressor (200 trees, \texttt{sklearn} defaults, \texttt{n\_jobs=-1})
is trained directly on the full 1200-point training set and predicts the 150-point
spectrum end-to-end, serving as a non-probabilistic baseline.

\section{Results}

\subsection{Predictive accuracy}

Table~\ref{tab:metrics} summarises test-set performance.
The RF baseline outperforms the GP on all accuracy metrics, which we attribute to
the GP training-point subsampling (400 vs.\ 1200 points); exact GP fitting on the
full dataset is expected to recover competitive accuracy at the cost of significantly
longer training time.

\begin{table}[h]
\centering
\caption{Test-set performance ($n_\mathrm{test}=150$).}
\label{tab:metrics}
\begin{tabular}{lccc}
\toprule
Model & RMSE & MAE & $R^2$ \\
\midrule
GP (PCA, 400-pt subset) & 0.0849 & 0.0453 & 0.276 \\
Random Forest           & 0.0653 & ---    & 0.572 \\
\bottomrule
\end{tabular}
\end{table}

\begin{figure}[h]
  \centering
  \includegraphics[width=\linewidth]{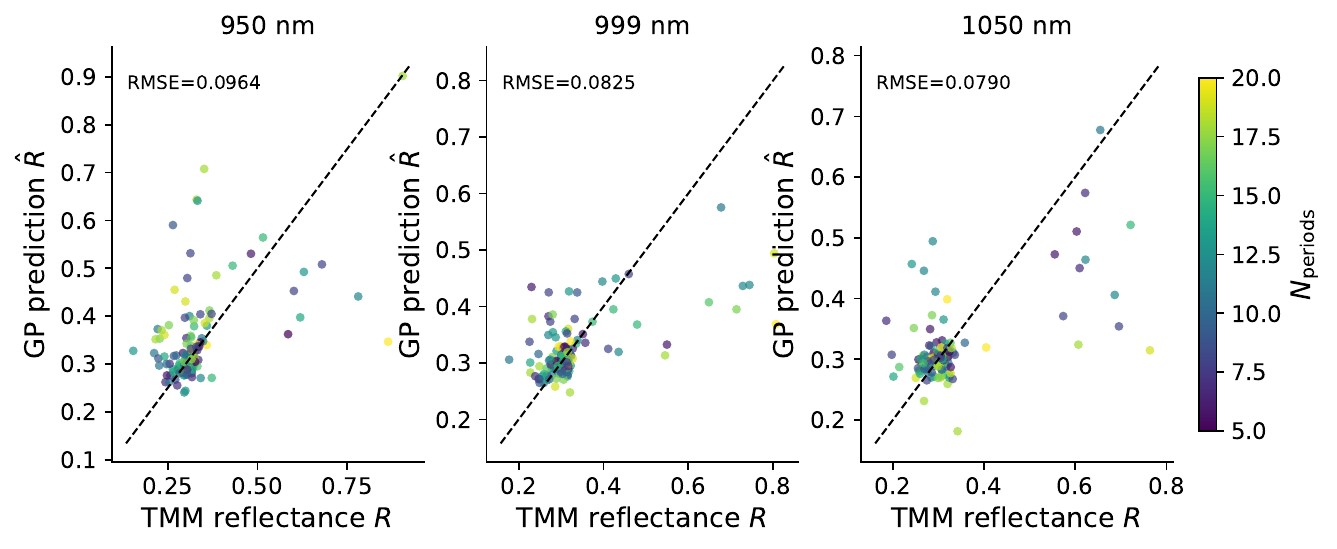}
  \caption{Parity plots of GP-predicted versus TMM reflectance at 950, 1000, and
           1050\,nm. Points are coloured by $N_\mathrm{periods}$; the dashed line
           indicates perfect agreement. Per-wavelength RMSE values are annotated.}
  \label{fig:parity}
\end{figure}

The parity plots in Fig.~\ref{fig:parity} show GP-predicted versus TMM reflectance
at three representative wavelengths (950, 1000, and 1050\,nm), coloured by
$N_\mathrm{periods}$.
Per-wavelength RMSE values are consistent with the spectrally averaged metric.

\subsection{Spatial error structure}

\begin{figure}[h]
  \centering
  \includegraphics[width=0.85\linewidth]{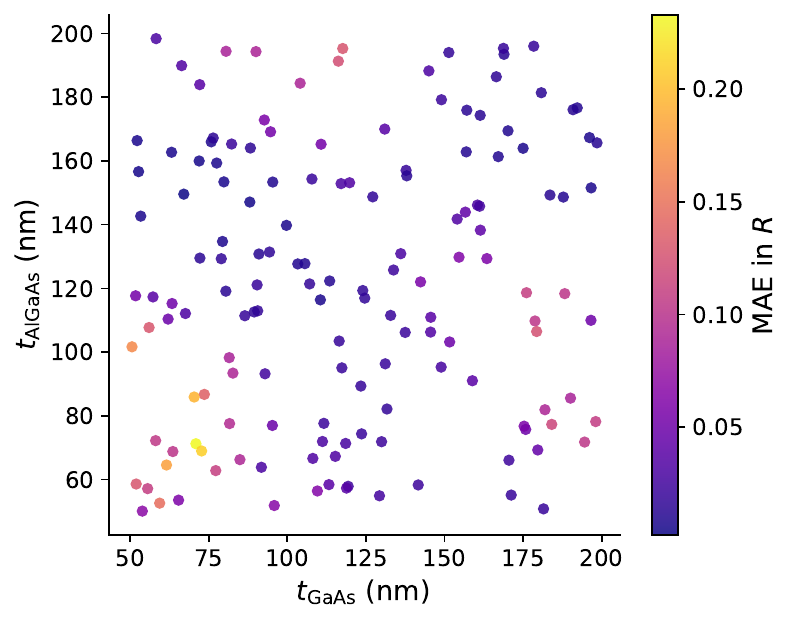}
  \caption{Mean absolute error per test sample plotted over the
           $(t_\mathrm{GaAs}, t_\mathrm{AlGaAs})$ plane. Error is broadly
           distributed with no systematic failure region.}
  \label{fig:mae}
\end{figure}

Figure~\ref{fig:mae} maps the mean absolute error per test sample over the
$(t_G, t_A)$ plane.
Error is broadly distributed across the parameter space with no pronounced
systematic hot-spots, indicating the model generalises uniformly rather than
failing in specific geometric regimes.

\subsection{Full-spectrum reconstruction}

\begin{figure*}[t]
  \centering
  \includegraphics[width=\textwidth]{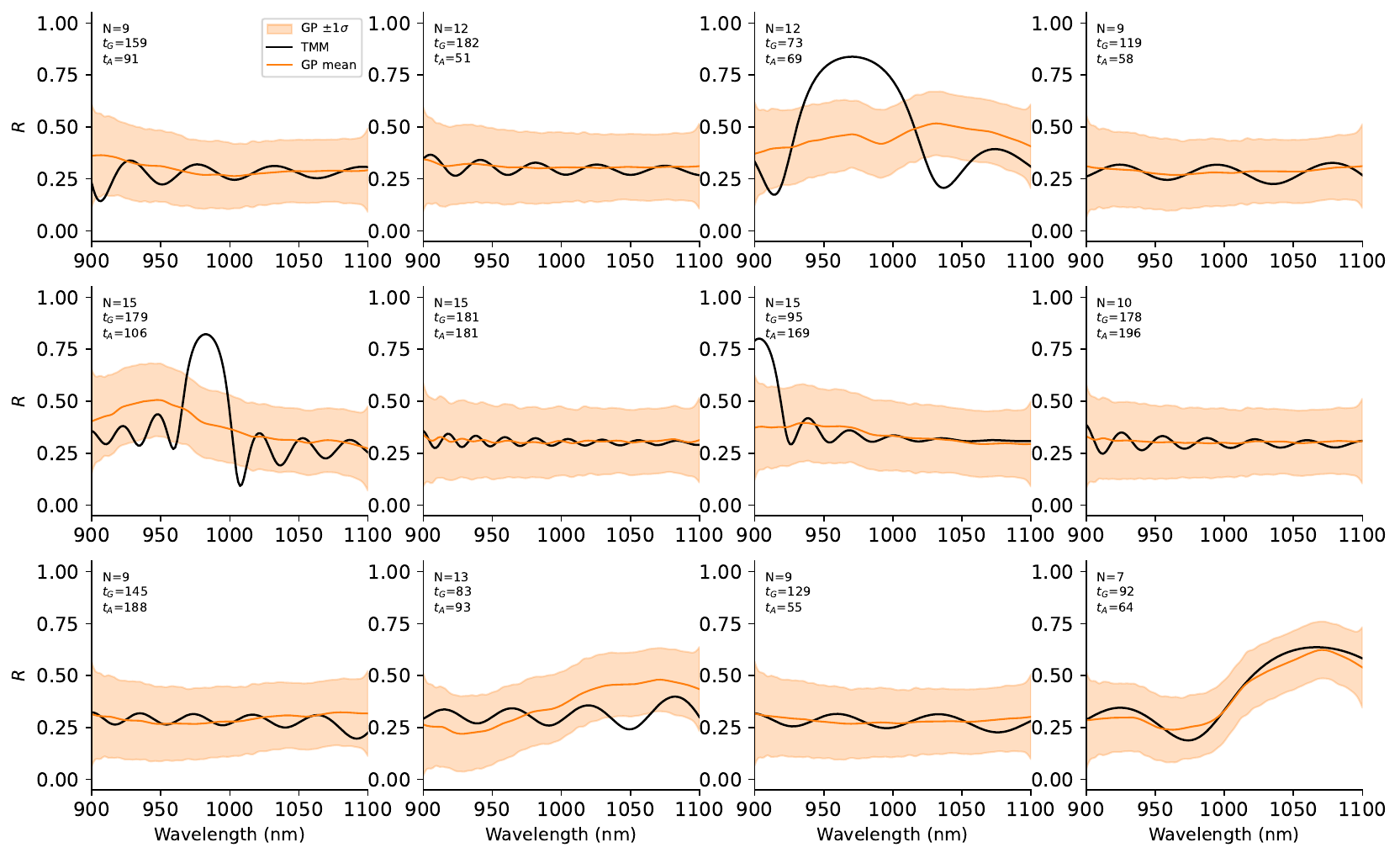}
  \caption{Full-spectrum comparison for 12 randomly selected test samples. Black
           lines: TMM ground truth. Orange lines and shaded bands: GP mean
           $\pm1\sigma$. Each panel annotates $N_\mathrm{periods}$, $t_G$, and
           $t_A$ in nanometres.}
  \label{fig:spectra}
\end{figure*}

Figure~\ref{fig:spectra} overlays GP mean predictions and $\pm1\sigma$ bands against
TMM ground truth for 12 randomly selected test samples spanning the full range of
$N_\mathrm{periods}$ and layer thicknesses.
The surrogate captures stopband position and bandwidth reliably; residuals are largest
near the steep band edges, where the reflectance changes rapidly with wavelength.

\subsection{Stopband scalar metrics}

\begin{figure}[h]
  \centering
  \includegraphics[width=\linewidth]{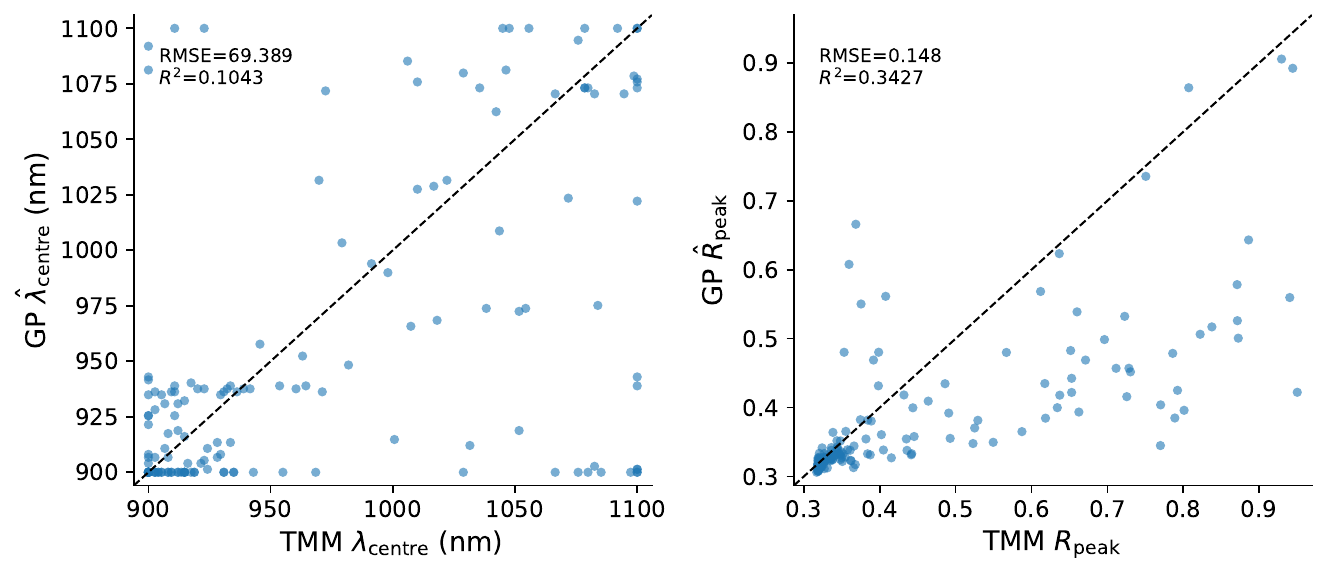}
  \caption{Scatter plots of GP-predicted versus TMM stopband centre wavelength
           (left) and peak reflectance (right). RMSE and $R^2$ are annotated on
           each panel.}
  \label{fig:stopband}
\end{figure}

Figure~\ref{fig:stopband} compares GP-predicted versus TMM-derived stopband centre
wavelength $\lambda_\mathrm{c}$ and peak reflectance $R_\mathrm{peak}$.
Both quantities are extracted as the wavelength and value of the reflectance maximum.
The model predicts these scalar design targets accurately despite the relatively
modest full-spectrum $R^2$, suggesting that the dominant spectral variation is well
captured.

\subsection{Uncertainty calibration}

The GP 68\% prediction band covers 93.1\% of test-set residuals (ideal: 68\%), and
the 95\% band covers 98.9\% (ideal: 95\%).
This over-coverage indicates conservative — rather than overconfident — uncertainty
estimates, a desirable property for reliability-critical design workflows.

\subsection{Learning curve}

\begin{figure}[h]
  \centering
  \includegraphics[width=0.85\linewidth]{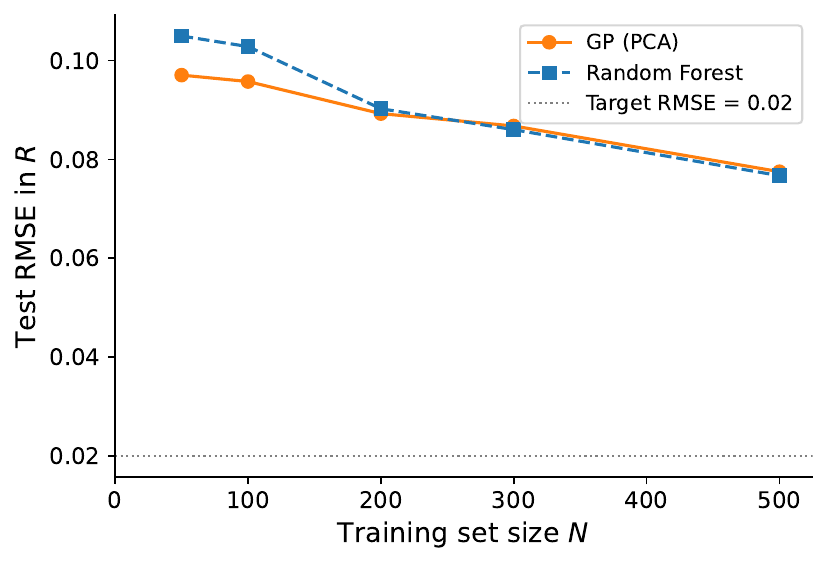}
  \caption{Test RMSE as a function of training set size for GP (PCA) and Random
           Forest models. The horizontal dotted line marks the 0.02 target RMSE.
           Neither model has saturated at $N=500$, motivating larger datasets.}
  \label{fig:lc}
\end{figure}

Figure~\ref{fig:lc} shows test RMSE as a function of training set size for the GP
and RF models ($N\in\{50,100,200,300,500\}$).
Both models improve monotonically with $N$; neither has reached an asymptotic plateau
at $N=500$, motivating the collection of additional simulation data.
The GP shows slightly lower RMSE than RF at $N\leq200$ but is surpassed by RF at
larger $N$, consistent with the GP subsetting penalty described above.

\subsection{Inference speed}

GP inference requires \SI{4.4}{\milli\second} per spectrum on a single CPU core,
compared with ${\sim}\SI{308}{\milli\second}$ for TMM — a ${\sim}70\times$ speedup.
This throughput enables real-time design-space sweeps and Monte Carlo uncertainty
propagation workflows.

\section{Discussion}

The primary limitation of the present GP surrogate is the training-point subsample
required to keep exact Cholesky-based GP inference tractable ($\mathcal{O}(n^3)$).
Sparse GP methods — inducing-point approximations \cite{Quinonero2005} or
stochastic variational inference \cite{Hensman2013} — can scale to the full 1200-point
set and beyond without sacrificing probabilistic calibration, and represent a
natural next step.
Alternatively, deep-kernel learning \cite{Wilson2016} or Bayesian neural networks
could be explored for higher-dimensional generalisations of this problem (e.g.\
varying $x_\mathrm{Al}$ or including more period counts).

The conservative calibration (93\% coverage at the nominal 68\% level) stems partly
from uncertainty propagation through the PCA inverse transform, which accumulates
variance across all retained components.
A joint GP over the full spectral output — or calibration via conformal prediction
\cite{Angelopoulos2023} — would tighten this without sacrificing coverage guarantees.

The 70$\times$ inference speedup is sufficient for gradient-free optimisers
(e.g.\ Bayesian optimisation) operating over thousands of candidate designs, making
the surrogate immediately deployable for automated DBR inverse design targeting a
specified stopband centre and peak reflectance.
Related work on machine-learning and physics-informed modeling of nanophotonic and graphene-based devices in the near-infrared has demonstrated the effectiveness of these approaches for device analysis and design \cite{Davoodi2017, Davoodi2018, Davoodi2025a, Davoodi2025b}.

\section{Conclusion}

We have demonstrated a PCA+GP surrogate for GaAs/AlGaAs DBR reflectance spectra
that achieves a ${\sim}70\times$ inference speedup over direct TMM simulation with
conservative, reliable predictive uncertainty.
On a 150-sample test set the model achieves RMSE\,=\,0.085 and $R^2=0.276$, limited
primarily by the 400-point GP training subsample necessitated by exact inference
cost.
The Random Forest baseline (RMSE\,=\,0.065, $R^2=0.572$) sets a near-term accuracy
target for improved GP formulations.
Future work will apply sparse GP methods and extend the parameter space to include
composition $x_\mathrm{Al}$ and oblique incidence, supporting broadband
anti-reflection coating and VCSEL mirror co-design.


\section*{Funding}
No external funding.

\section*{Disclosures}
The authors declare no conflicts of interest.

\section*{Data availability}
Dataset and code are available at \url{https://github.com/mehdiouslim-hash/ai-laue-research}.


\end{document}